\documentclass[review]{elsarticle}

\usepackage{hyperref,graphicx,amsmath,amssymb,bm,lscape,multirow}

\journal{Journal of \LaTeX\ Templates}

\begin{document}

\begin{frontmatter}

\title{Complete Glauber calculations for proton-nucleus inelastic cross
  sections}

\author{S. Hatakeyama and W. Horiuchi*}
\address{Department of Physics, Hokkaido University, Sapporo 060-0810, Japan}
\cortext[mycorrespondingauthor]{Corresponding author}



\begin{abstract}
  We perform a parameter-free calculation
  for the high-energy proton-nucleus
  scattering based on the Glauber theory.
  A complete evaluation of the so-called Glauber amplitude is made
  by using the factorization of the single-particle wave functions.
  The multiple-scattering or multistep processes
  are fully taken into account within the Glauber theory.
  We demonstrate that proton-$^{12}$C, $^{20}$Ne, and $^{28}$Si elastic
  and inelastic scattering ($J^\pi=0^+ \to 2^+$ and $0^+\to 4^+$)
  processes are very well described
  in a wide range of the incident energies from
  $\sim 50$ MeV to $\sim 1$ GeV.
  We evaluate the validity of a simple one-step approximation
  and find that the approximation works fairly well for
  the inelastic $0^+ \to 2^+$ processes but not for $0^+\to 4^+$ where
  the multistep processes become more important.
  As an application, we quantify the difference between the total reaction
  and interaction cross sections of proton-$^{12}$C, $^{20}$Ne, and $^{28}$Si
  collisions.
\end{abstract}


\end{frontmatter}


\section{Introduction}

Recent major upgrades in radioactive beam facilities provide
the platform to study the exotic phenomena in
the unstable nuclei far from the stability line.
The understanding of the role of the excess neutrons in isotopic chains
has been deepened through the studies of the nuclear excitations using exotic
radioactive-ion beams,
for example, a systematic measurement of quadrupole transition strengths
has shown anomalous structure changes
due to neutron excess in the neutron-rich carbon isotopes
~\cite{Ong08, Wiedeking08, Petri11, Voss12}.

Since short-lived nuclei cannot be used as a target
nucleus, the nuclear direct reactions in the inverse kinematics
have often been utilized as a tool to study the structure
of such nuclei. A proton, which is the simplest probe,
has often been used to populate the excited states of nuclei.
Thanks to high-intensity radioactive beams,
the proton inelastic scattering cross section measurements
of the short-lived nuclei have become possible
with use of the inverse kinematics.~\cite{Kraus94, Aoi12, Tanaka17}.
In contrast to electron and photon scattering,
both proton and neutron parts of the projectile nucleus can
directly be excited through the proton inelastic scattering processes.
This is advantageous for studying
the detailed structure of the neutron-rich nuclei
where the neutron excitations are expected to be dominant.
Here we focus on the proton-nucleus inelastic scattering
at about 50 to the several hundred MeV where
the measurements have often been made.
This high-energy region is beneficial
for a theoretical description
as the reaction mechanism is much simpler than
the low-energy region in which the complicated channel coupling
effects should be taken into account~\cite{Matsumoto17}.

Towards the future measurements of the inelastic scattering
cross sections for unstable nuclei,
in this paper, we develop a parameter-free reaction theory
based on the Glauber theory~\cite{Glauber}
and test it in comparison to the available experimental data.
The Glauber theory is one of the most widely accepted methods to describe
the nuclear reactions at high incident energies.
We evaluate proton-nucleus inelastic scattering cross sections
following the original Glauber theory which includes
all multiple-scattering or multistep processes within
the eikonal and adiabatic approximations.
According to the original formulation of the Glauber theory,
the inputs to the theory are wave functions (not one-body densities)
of the colliding nuclei and the so-called profile function
parametrized based on the total nucleon-nucleon cross section.
Therefore, the theory includes no adjustable parameter.
Most complicated part of the computation
is the evaluation of the so-called Glauber amplitude
involving multidimensional integration, which is in general difficult,
and often approximate treatment has been made to avoid that difficulty.
By introducing appropriate approximations,
the theory successfully reproduced the observed cross sections
of the unstable nuclei and revealed the evolution of the
nuclear deformation in the neutron-rich
isotopes~\cite{Horiuchi12,Horiuchi15}. However,
the complete and approximated Glauber amplitudes significantly deviate
in case of halo nuclei where the nuclear surface
is very much extended~\cite{Ogawa92, Al-Khalili96, Al-Khalili96b, Nagahisa18}.
Since the inelastic scattering occurs mainly around the nuclear surface,
the complete evaluation of the Glauber amplitude which
includes all the multistep processes in the Glauber theory
will be necessary for a more reliable description
of the scattering processes.

The purpose of this paper is to establish
a reliable microscopic framework
following the original Glauber theory
towards future proton-nucleus inelastic
cross section measurements involving the exotic nuclei.
We remark that Ref.~\cite{Abgrall76} reported
the complete Glauber calculations for
proton-$^{12}$C inelastic scattering cross sections
and successfully reproduced the cross sections at $\sim$1\,GeV.
However, the form of the wave function they used is limited to
an analytically integrable form such as harmonic-oscillator wave functions
in which applications to heavier nuclei
as well as extension to more general wave function is difficult.
In the present study, we extend this approach
in order to use more general forms of the wave functions.
To demonstrate the power of this approach,
we systematically analyze the inelastic scattering
cross sections for well known nuclei $^{12}$C, $^{20}$Ne, and $^{28}$Si,
and compare them with the available experimental data.

The paper is organized as follows. In Sec.~\ref{models.sec}, 
we briefly explain the Glauber theory to describe
the nuclear elastic and inelastic processes.
In Sec.~\ref{formulation.sec}, the formulation
to compute these cross sections
is given based on the Glauber multiple-scattering theory.
Sec.~\ref{Gamp.sec} explains how we evaluate the complete Glauber amplitude
for the elastic and inelastic scattering cross section calculations.
For later use, approximate formulation to evaluate the Glauber amplitude
is given in Sec.~\ref{approx.sec}. 
This theory will be tested for the evaluation of
the elastic and inelastic cross sections of 
$^{12}$C, $^{20}$Ne, and $^{28}$Si.
Though the theory can use any type of
the single-particle wave functions,
we, however, employ deformed harmonic-oscillator
wave functions for the sake of simplicity
which are defined in Sec.~\ref{wf.sec}. 
Section~\ref{results.sec} discusses our results of
the elastic and inelastic scattering cross sections.
We show the physical properties of our wave functions in
Sec.~\ref{wfres.sec}.
Section~\ref{csres.sec} compares the theoretical elastic and inelastic
scattering cross sections with the available cross section data.
The approximate methods are also tested in this section
in order to quantify the importance of the multiple-scattering
or multistep processes which have often been neglected.
The structure of $^{28}$Si is discussed through
a systematic analysis of the inelastic scattering cross sections.
The energy dependence of the inelastic scattering processes
is discussed in Sec.~\ref{inel.sec}.
As an application of this theory, in Sec.~\ref{sigmai.sec},
we evaluate difference between the total reaction
and interaction cross sections as they impacts on
the accuracy of the radius extraction from the measured
interaction cross section. A summary is given in Sec.~\ref{summary.sec}.
More details about the evaluation of the Glauber amplitude are described
in Appendices A and B.

\section{Theoretical models}
\label{models.sec}

The Glauber theory~\cite{Glauber} is a powerful tool to describe
the scattering processes in high-energy nucleus-nucleus collisions.
In this section, we summarize how the scattering cross sections
are evaluated with the Glauber theory.
The Glauber amplitude is a key to the calculation of all the cross sections.
Here we explain a procedure to compute it
for proton-nucleus scattering.

\subsection{Inelastic scattering cross sections within the Glauber theory}
\label{formulation.sec}

We consider the normal kinematics throughout this paper
in which a high-energy proton is bombarded on a target nucleus
for the sake of convenience,
and assume that this incoming proton is not polarized.
In the Glauber theory, the final state wave function
of a proton and mass number $A$ system, $\Phi_f$,
is greatly simplified with the help of the adiabatic
and eikonal approximations as~\cite{Glauber}
\begin{align}
\Phi_f= \left[\prod_{j=1}^Ae^{i\chi_{pN}({\bm{b}-\hat{\bm{s}}_j})}\right]\Phi_0,
\end{align}
in which $\Phi_f$ is expressed by the product
of the initial-(ground-)state wave function, $\Phi_0$, and
the product of the proton-nucleon ($pN; N=p$ or $n$ for proton
or neutron) phase-shift functions $e^{i\chi_{pN}(\bm{b}-\hat{\bm{s}}_j)}$
with $\hat{\bm{s}}_j$ being the two-dimensional single-particle
coordinate operator of the $j$th nucleon
perpendicular to the beam direction $z$.
We conveniently define the Glauber multiple-scattering operator as
\begin{align}
\prod_{j=1}^Ae^{i\chi_{pN}({\bm{b}-\hat{\bm{s}}_j})}
=\prod_{j=1}^A\left[
   1-\Gamma_{pN}(\bm{b}-\hat{\bm{s}}_j)\right]
=\prod_{j=1}^A \mathcal{O}_j(\bm{b})
\end{align}
with the $pN$ profile function, $\Gamma_{pN}(\bm{b})$,
which is usually parametrized as~\cite{Ray79}
\begin{align}
  \Gamma_{pN}(\bm{b})=\frac{1-i\alpha_{pN}}{4\pi\beta_{pN}}\sigma_{pN}^{\rm tot}
  \exp\left[-\frac{\bm{b}^2}{2\beta_{pN}}\right],
\end{align}
where $\sigma_{pN}^{\rm tot}$, $\alpha_{pN}$, and $\beta_{pN}$
are the total $pN$ cross section, the ratio between
the real and imaginary parts of the scattering amplitude
at the forward angle, and the slope parameter, respectively.
These parameter sets for various incident energies are
taken from Ref.~\cite{Ibrahim08}.
The validity of the profile function has been confirmed
in a number of examples, not only
for nucleon-nucleus scattering
but also nucleus-nucleus scattering~\cite{Ibrahim09,Horiuchi10,Horiuchi12,Suzuki16,Horiuchi16,Nagahisa18},
and thus the profile function in Ref.~\cite{Ibrahim08}
can be regarded as one optimal choice, although
there are some ambiguity due to the experimental uncertainty,
especially at the low incident energies~\cite{Ibrahim08}.
In the incident energies below the pion production threshold,
the nucleon-nucleon elastic scattering differential cross sections
obtained from a realistic nucleon-nucleon interaction 
will be useful to reduce the uncertainty
of the profile function fixed by the data fitting.
This is interesting and worth investigating in the future.

The scattering amplitude from the initial ground state $(\alpha=0)$
to the final state labeled with $\alpha$ 
can be calculated by~\cite{Glauber,Suzuki03}
\begin{align}
  f_{\alpha}(q)&=\frac{k}{2\pi i}\int d\bm{b}\,
  e^{i\bm{q}\cdot\bm{b}}
  \left[\delta_{\alpha,0}-\left<\Phi_\alpha\right|
\prod_{j=1}^A \mathcal{O}_j(\bm{b})
  \left|\Phi_0\right>\right],
\label{samp.eq}
\end{align}
where $k$ is the wave number in the relativistic kinematics,
$\bm{q}$ is the momentum transfer vector being
$|\bm{q}|=q=2k\sin(\theta/2)$ with the scattering angle $\theta$
in the center-of-mass (cm) system.
The orthogonormality relation
$\left<\Phi_\alpha|\Phi_0\right>=\delta_{\alpha,0}$ is used in this derivation.
In Appendix A, we give more details about the evaluation
of Eq.~(\ref{samp.eq}).

The elastic $(\alpha=0)$ and inelastic $(\alpha\neq 0)$
scattering differential cross sections can be
evaluated by
\begin{align}
  \frac{d\sigma_{\alpha}}{d\Omega}=\frac{v_\alpha}{v_0}|f_{\alpha}(q)|^2,
\end{align}
where $v_0$ and $v_\alpha$ are the velocities of
the initial-incoming and final-outgoing waves, respectively.
In the adiabatic approximation, $v_\alpha/v_0$ is unity.
This is reasonable when the beam energy
is high enough as compared to the excitation energy of the nucleus. 
The inelastic scattering cross section 
can directly be obtained by integrating the differential cross sections
over the scattering angles with $\alpha\neq 0$
\begin{align}
  \sigma_{\alpha}&=\int d\Omega\,
  \frac{d\sigma_{\alpha}}{d\Omega}
  =\int d\bm{b}\,|\left<\Phi_\alpha\right|
\prod_{j=1}^A \mathcal{O}_j(\bm{b})
  \left|\Phi_{0}\right>|^2.
\end{align}
It can be rewritten in terms of the so-called Glauber amplitude
for the inelastic processes
\begin{align}
  \mathcal{T}_{\alpha} (J_0M_0\to J_{\alpha}M_{\alpha};\bm{b})
  =\left<\Phi_{\alpha; J_{\alpha}M_{\alpha}}\right|
\prod_{j=1}^A \mathcal{O}_j(\bm{b})
\left|\Phi_{0; J_0M_0}\right>,
\label{gainel.eq}
\end{align}
where $J_0M_0$ $(J_\alpha M_\alpha)$ is the
the initial (final) angular momentum and its projection.
The inelastic scattering cross section
is evaluated by the expression
\begin{align}
  \sigma_{\rm \alpha}=\frac{1}{2J_0+1}\sum_{M_0,M_\alpha}
  \int d\bm{b}\, \left|\mathcal{T}_{\alpha} (J_0M_0\to J_\alpha M_\alpha;\bm{b})\right|^2.
\end{align}

\subsection{Evaluation of the complete Glauber amplitude
  for the inelastic scattering}
\label{Gamp.sec}

Evaluation of the Glauber amplitude of Eq.~(\ref{gainel.eq})
requires in general the tedious computations
as one has to evaluate the $A$-fold multidimensional integration.
A Monte Carlo technique was successfully applied to
evaluate the multidimensional integration~\cite{Varga02,Nagahisa18}.
However, it cannot be applied to
the inelastic scattering problem because
the initial and final states are orthogonal in which
the guiding function $\Phi_\alpha^*\Phi_0$ for the Metropolis
algorithm~\cite{Metropolis53} is no longer positive definite.

In the present work, we take another approach
based on the idea presented in Ref.~\cite{Bassel68}.
With use of a Slater determinant wave function,
the Glauber amplitude is factorized and 
its multidimensional integration
is reduced to the three-dimensional one
on the single-particle coordinate,
which can simply be evaluated by a standard numerical
integration technique, e.g., the trapezoidal rule and the Gaussian quadrature.
This factorization technique has been successfully applied to
realistic proton-nucleus elastic scattering
of various nuclear systems~\cite{Ibrahim09,Hatakeyama14,Hatakeyama15}.
In order to apply this method to the proton-nucleus
inelastic scattering computation,
here we extend the expression in order to use the wave function
expressed by multi-Slater determinants.
An earlier study was done for the proton-$^{12}$C inelastic scattering
with a specific form of the wave function~\cite{Abgrall76}.
Here we generalize it towards the application of using the realistic
nuclear wave functions such as from the shell model, the mean-field model
as well as the antisymmetrized- and fermionic-molecular dynamics models
~\cite{AMD, FMD1, FMD2}.

We assume that the total wave function is expressed by
a superposition of the antisymmetrized product
of the single-particle wave functions as
\begin{align}
\Phi_\alpha=\sum_{i}
C_i^{(\alpha)}\mathcal{A}\left\{\prod_{j=1}^A\varphi_{i_j}^{(\alpha)}\right\},
\qquad \varphi_{i_j}^{(\alpha)}=\phi_{i_j}^{(\alpha)}\chi_{i_j}^{(\alpha)}
\xi_{i_j}^{(\alpha)},
\label{slwf.eq}
\end{align}
where $\mathcal{A}$ is the antisymmetrizer,
and $\phi_{i_j}^{(\alpha)}$, $\chi_{i_j}^{(\alpha)}$, and $\xi_{i_j}^{(\alpha)}$
denote the $j$th single-particle orbital, spin, and isospin wave functions
belonging to the state $\alpha$, respectively.
The Glauber amplitude of Eq.~(\ref{gainel.eq}) is written
explicitly using the definition (\ref{slwf.eq}) as
\begin{align}
\mathcal{T}_\alpha=\sum_{i,k}C_{i}^{(\alpha)}C_{k}^{(0)}
      {\rm det}\left\{\left<\varphi_{i_j}^{(\alpha)}\right|
    \mathcal{O}_{j}(\bm{b})
    \left|\varphi_{k_l}^{(0)}\right>\right\}
\qquad (j,l=1,\dots, A),
    \label{gaex.eq}
\end{align}
in which the multidimensional integration of Eq.~(\ref{gainel.eq})
is reduced to a calculable 3-fold integration in the orbital part.
We note that the single-particle wave function in the above equation
are not necessarily to be orthogonal with each other.
We describe more details how to evaluate
Eq.~(\ref{gaex.eq}) in Appendix B.

\subsection{Approximations of the Glauber amplitude}
\label{approx.sec}

In this study, we fully include the multiple-scattering or multistep processes
within the Glauber theory.
In order to see these effects in the cross sections,
we compare the cross sections obtained with some approximate methods.
The optical-limit approximation (OLA) has widely been applied
as it only requires the nuclear density distribution of the target nucleus.
The OLA is derived by the leading order
of the cumulant expansion of the Glauber amplitude as~\cite{Glauber,Suzuki03}
\begin{align}
  \mathcal{T}_0(\bm{b})\simeq\mathcal{T}_0^{\rm OLA}(\bm{b})=
  \exp\left(-\sum_{N=n,p}\int d\bm{r}\,\rho_{00}^{(N)}(\bm{r})\Gamma_{pN}(\bm{b}-\bm{s})
  \right)
\label{OLA.eq}
\end{align}
with $\bm{r}=(\bm{s},z)$,
where $\rho_{00}^{(N)}$ is the one-body density of the target nucleus
for proton or neutron, which is more generally defined by
\begin{align}
\rho_{\alpha 0}^{(N)}(\bm{r})=\sum_{j \in N}
  \left<\Psi_\alpha\right|\delta(\hat{\bm{r}}_j-\bm{r})\left|\Phi_{0}\right>,
\end{align}
where $\hat{\bm{r}}_j$ is the single-particle coordinate operator
of the $j$th nucleon.
Since the cumulant expansion is a series expansion with respect
to the moment of the function, it cannot be applied directly
for the inelastic scattering due to the orthogonality of
the initial and final state wave functions.
For the elastic scattering,
by assuming the factorization of the $A$-body density
and taking only one-step contribution~\cite{Abgrall76},
we get
\begin{align}
  \mathcal{T}_0(\bm{b})\simeq\bar{\mathcal{T}}_0(\bm{b})=
  \left[1-\bar{\Gamma}_{00}^{(p)}(\bm{b})\right]^Z
  \left[1-\bar{\Gamma}_{00}^{(n)}(\bm{b})\right]^{A-Z},
\label{DWIA-1.eq}
\end{align}
where $Z$ is the atomic number of the target nucleus with
\begin{align}
  \bar{\Gamma}_{\alpha 0}^{(p)}(\bm{b})
  &=\frac{1}{Z}\int d\bm{r}\,\rho_{\alpha0}^{(p)}(\bm{r})
  \Gamma_{pp}(\bm{b}-\bm{s}),\\
\bar{\Gamma}_{\alpha 0}^{(n)}(\bm{b})
&=\frac{1}{A-Z}\int d\bm{r}\,\rho_{\alpha0}^{(n)}(\bm{r})
\Gamma_{pn}(\bm{b}-\bm{s}).
\end{align}
 The same assumption is also applied
to the inelastic scattering case $(\alpha\neq 0)$ 
\begin{align}
  \bar{\mathcal{T}}_\alpha^{(p)}(\bm{b})&=  
  A\left[1-\bar{\Gamma}_{00}^{(p)}(\bm{b})\right]^{Z-1}
  \left[1-\bar{\Gamma}_{00}^{(n)}(\bm{b})\right]^{A-Z}
  \bar{\Gamma}_{\alpha 0}^{(p)}(\bm{b})\\
  \bar{\mathcal{T}}_\alpha^{(n)}(\bm{b})&=  
  A\left[1-\bar{\Gamma}_{00}^{(p)}(\bm{b})\right]^{Z}  
  \left[1-\bar{\Gamma}_{00}^{(n)}(\bm{b})\right]^{A-Z-1}
  \bar{\Gamma}_{\alpha 0}^{(n)}(\bm{b})
\label{DWIA-2.eq}
\end{align}
for proton and neutron excitations, respectively.
To get an approximated Glauber amplitude,
we take an average of the proton and neutron amplitudes
as
\begin{align}
  \mathcal{T}_\alpha(\bm{b})\simeq
  \bar{\mathcal{T}}_\alpha(\bm{b})
  =\frac{\bar{\mathcal{T}}_\alpha^{(p)}(\bm{b})+
     \bar{\mathcal{T}}_\alpha^{(n)}(\bm{b})}{2} \qquad (\alpha\neq 0).
\end{align}
  This is nothing but the expression
  of the eikonal version
  of the distorted-wave-impulse approximation (DWIA)~\cite{Saudinos74,Alexander74}.

\subsection{Wave function}
\label{wf.sec}

As inputs to the theory, we need wave functions
of the initial (ground) and final (excited) states of the target nucleus.
For the sake of simplicity, in this paper,
we consider the ground and excited states are respectively
generated by the angular momentum projection of
a single-Slater determinant intrinsic wave function.
The wave function in the laboratory frame
with the total spin $J$ and its projection $M$ is
obtained by the angular momentum projection
\begin{align}
  \Phi_{JM}=\mathcal{N}_{MK}^{J}
  \int d\omega\,[\mathcal{D}_{MK}^{J}(\omega)]^*
  \hat{\mathcal{R}}(\omega)\Phi^{\rm int}_{K},
\end{align}
where $\mathcal{N}_{MK}^{J}$ is a normalization constant,
$\mathcal{D}_{MK}^{J}(\omega)$ is the Wigner D-function,
and $\hat{\mathcal{R}}(\omega)$
is the rotation operator
with respect to the Euler angles $\omega=(\theta_1,\theta_2,\theta_3)$,
which acts on the orbital and spin coordinates
of the intrinsic wave function.
Note that the resulting total wave function is 
expressed with a multi-Slater determinant. 

To make the calculation simpler,
the intrinsic total wave function with the projection
on the symmetry axis $z$ is assumed as the product of
the axially-symmetric deformed
harmonic-oscillator (DHO) single-particle wave functions 
\begin{align}
  \Phi^{\rm int}_{K}=\mathcal{A} \left\{\prod_{j=1}^A\phi_{{\bar{N}}_j{n_{z}}_j\Lambda_j}(\bm{r}_j)\chi_{\frac{1}{2}m_j}\xi_{\frac{1}{2}{\bar{m}}_j}\right\},
  \label{intwf.eq}
\end{align}
where $\phi$, $\chi$, and $\xi$ respectively denote
the orbital, spin, and isospin wave functions; and
$\bar{N}_j$, ${n_{z}}_j$, $\Lambda_j$,
$m_j$, and $\bar{m}_j$ are the total quantum number,
the quantum number of the symmetry axis, the projection of the
orbital angular momentum onto the symmetry axis,
the intrinsic spin, and the isospin of the $j$th nucleon, respectively.
In this paper, we take up $J^\pi=0^+, 2^+$ and $4^+$ states
belonging to the ground-state rotational band of the three
closed shell ($Z=6$, 10, 14) nuclei, $^{12}$C, $^{20}$Ne, and $^{28}$Si.
In the axially-symmetric DHO shell model,
these positive-parity states correspond to $K=0$.
This model works well for these nuclei
as was shown in Ref.~\cite{Abgrall72}.

In the actual computations, the rotation with respect to $\theta_3$
is redundant in the axially-symmetric case.
To ensure 3 digit accuracy in physical quantities of the wave function,
we take 20 points respectively for $\theta_1$ and $\theta_2$, 
which results in a superposition of 400 Slater determinants
at each angular mesh point
whose weight factors [$C_i^{(\alpha)}$ in Eq.~(\ref{slwf.eq})] are
determined through the Wigner D-function.

\section{Comparison of the theory and experiment}
\label{results.sec}
\subsection{Properties of the wave functions}
\label{wfres.sec}

Configurations of the wave functions taken into account
are summarized and listed in Table~\ref{conf.tab}.
We assume that the proton and neutron configurations are the same.
We remark that the single-particle energy of the DHO wave function 
$\hbar\omega_0\left[\bar{N}+\frac{3}{2}+(\bar{N}-3n_z)\frac{\epsilon}{3}\right]$
is expressed by two parameters, the averaged oscillator frequency,
$\omega_0=(2\omega_\perp+\omega_z)/3$,
and the ratio of difference between the oscillator frequencies of
the symmetric and the other axes to the averaged oscillator frequency,
$\epsilon=3(\omega_\perp-\omega_z)/(2\omega_\perp+\omega_z)$,
with the oscillator frequency of the symmetric axis $z$, $\omega_z$,
and the one perpendicular to $z$, $\omega_\perp$.
Obviously, the wave functions are oblate for $^{12}$C
and prolate for $^{20}$Ne. For $^{28}$Si, the oblate and
prolate configurations can be assumed.
The ground $0^+$, and excited 2$^+$, 4$^+$ states
are generated by the angular momentum projection.
The two parameters, $\omega_0$ and $\epsilon$,
determine the characteristics of the
wave function, that is, the nuclear size and the degree of deformation,
and are fixed in such a way so as to reproduce the measured charge radius
and the reduced electric-quadrupole transition probability
simultaneously with the angular momentum projected total wave function
for the $0^+$ and $2^+$ states.

\begin{table}[b]
  \caption{Configurations 
    of the intrinsic wave functions of $^{12}$C, $^{20}$Ne,
    and $^{28}$Si expressed in the quantum numbers
    of the axially-symmetric deformed
    harmonic-oscillator wave function $[\bar{N}n_z|\Lambda|]$
    with two oscillator parameters, $\omega_0$ and $\epsilon$.
  See text for details.}
\label{conf.tab}
\centering
\begin{tabular}{cccccc}
\hline
Nucleus &&  Configurations&&$\omega_0$ ($c/{\rm fm}$)&$\epsilon$\\ 
\hline
$^{12}$C  && $[000]^2[101]^4$&&0.0953  &$-$0.594 \\
$^{20}$Ne && $[000]^2[110]^2[101]^4[220]^2$&&0.0719 &0.484 \\
$^{28}$Si (oblate)&&$[000]^2[101]^4[110]^2[200]^2[202]^4$&&0.0675 &$-$0.250 \\
$^{28}$Si (prolate)&&$[000]^2[110]^2[101]^4[220]^2[211]^4$&&0.0666 &0.152 \\
\hline
\end{tabular}
\end{table}

We define some physical quantities
which are useful to show the properties of the wave functions.
The root-mean-square (rms) point-proton radius and the reduced electric
transition probabilities with the multipolarity $\lambda$
are respectively calculated by
\begin{align}
  \left<r_p^2\right>
  =\frac{1}{Z}\left<\Phi_{00}\right|\sum_{j\in p}^{Z}\hat{r}_j^2\left|\Phi_{00}\right>
\end{align}
and
\begin{align}
  B(E\lambda;J_i \to J_f)
  =\frac{1}{2J_i+1}\sum_{M_f,M_i,\mu}
  \left|\left<\Phi_{J_fM_f}\right|\sum_{j\in p}^Z \hat{r}_j^\lambda Y_{\lambda\mu}
  (\hat{\Omega}_j)\left|\Phi_{J_iM_i}\right>\right|^2.
\end{align}
As a measure of quadrupole deformation
it is useful to calculate the quadrupole 
deformation parameter of the intrinsic wave function defined by
\begin{align}
  \beta_{2}=\sqrt{\frac{5}{\pi}}\frac{3\left<z^2\right>-\left<r^2\right>}
       {\left<r^2\right>},
\end{align}
where $r^2=x^2+y^2+z^2$ with the symmetry axis $z$
and the axis $x(=y)$ perpendicular to $z$,
and $\left<\dots\right>$ denotes
the expectation value with the intrinsic
wave function of Eq.~(\ref{intwf.eq})
as $\left<\Phi^{\rm int}_{K}\right|\dots\left|\Phi^{\rm int}_{K}\right>$.

Table~\ref{wf.tab} lists the physical quantities
obtained with the wave functions of $^{12}$C , $^{20}$Ne, and $^{28}$Si.
As one can see from the table, we find
good matching for the $\sqrt{\left<r^2_p\right>}$ and $B(E2)$ values
within the DHO models for $^{12}$C, $^{20}$Ne, and $^{28}$Si,
resulting in considerably quadrupole-deformed total wave functions.
We remark that the $B(E2)$ value is only determined by the
absolute value of the quadrupole deformation parameter $|\beta_2|$.
In fact, the oblate and prolate wave functions of $^{28}$Si
give the same $|\beta_2|$ and $B(E2)$ values.

It should be noted that
all the physical quantities are measured from the origin
of the single-particle coordinate in the present paper,
which include the cm contribution.
For spherical HO wave functions,
we can exactly remove the cm wave function from
the total wave function~\cite{Ibrahim09},
whereas it is not trivial in the case of the DHO wave functions.
We, however, assume the origin of the coordinate
as the cm of the system in the present paper.
To keep the consistency in the calculations,
we use the same wave function to the cross section calculations as well.
Since the parameters in the wave function are fixed
so as to reproduce some physical quantities
without the cm correction, the cm effects are somewhat renormalized into
those parameters through the fit.
We confirm in $^{16}$O case with the spherical HO wave function
that the elastic scattering cross sections
at the forward angles do not change
with the cm corrected and uncorrected  (refitted) wave functions.
The little difference appears at larger scattering angles
due to the correction factor
$\exp(-\nu^2q^2/4A)$~\cite{Ibrahim09},
with $\nu$ being the size parameter of the HO wave function, multiplied to
 the elastic scattering differential cross sections.
 
\begin{table}[!h]
  \caption{Properties of the wave functions of $^{12}$C, $^{20}$Ne,
    and $^{28}$Si. Point-proton radii
    are extracted from the charge radius measurements~\cite{Angeli13}
    Experimental data of
    the reduced electric-quadrupole
    transition probabilities, $B(E2;2^+\to 0^+)$,
    are taken from Ref.~\cite{BE2} and averaged.
  } 
\label{wf.tab}
\centering
\begin{tabular}{ccccccccc}
\hline
        &&\multicolumn{2}{c}{$\sqrt{\left<r_p^2\right>}$ (fm)}&&
\multicolumn{2}{c}{$B(E2)$ ($e^2$fm$^4$)}&\\
\cline{3-4}\cline{6-7}
Nucleus &&Theo. & Expt.&& Theo.& Expt.& $\beta_2$\\
\hline
$^{12}$C  &&2.33 &2.327$\pm 0.009$&&8.24 &8.30$\pm 1.19$&$-$0.443\\ 
$^{20}$Ne &&2.89 &2.889$\pm 0.009$&&77.8 &77.7$\pm 12.2$&0.572\\ 
$^{28}$Si (oblate)&&3.01 &\multirow{2}{*}{3.010$\pm 0.009$}&&67.1 &\multirow{2}{*}{66.9$\pm 10.2$}&$-$0.339\\
$^{28}$Si (prolate)&&3.01&&&66.9 &&0.336\\ 
\hline
\end{tabular}
\end{table}

\subsection{Elastic and inelastic scattering differential cross sections}
\label{csres.sec}

\begin{figure}[!h]
\centering\includegraphics[width=\linewidth]{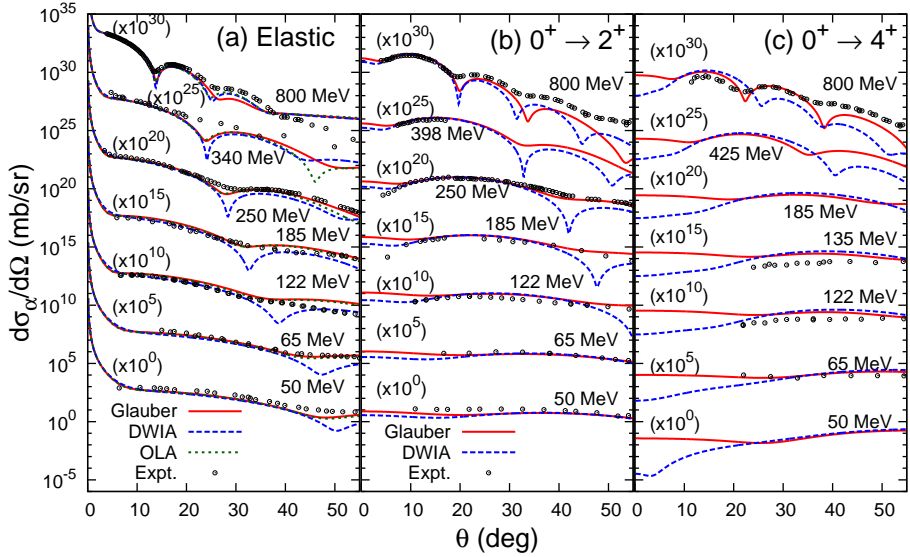}
\caption{(a) Elastic and (b) inelastic scattering differential cross sections
  of $0^+ \to 2^+$ and (c) $0^+ \to 4^+$
  for proton-$^{12}$C scattering. The results with
  the optical-limit approximation (OLA)
  and the eikonal-distorted-wave-impulse approximation (DWIA)
  are also plotted for comparison.
Experimental data are taken from Refs.~
\cite{RBS71,ISNSO87,KOKHS85,MSJH83,CADMF80,IJH79,MSAMJ88,RBLM52,BHBMG81}
for the elastic scattering,
Refs.~\cite{FBSG67,CBSD70,KOKHS85,CADMF80,HBM83,IJH79,MSAMJ88,JGSBC94,BCLHR78}
for the $0^+\to 2^+$ inelastic scattering, and Refs.~
\cite{KOKHS85,CADMF80,BCHSS83,BCLHR78} for the $0^+\to 4^+$ inelastic scattering. Experimental error bars are omitted since they are small.}
\label{12Cdiff.fig}
\end{figure}

Figure~\ref{12Cdiff.fig} plots the elastic and inelastic differential scattering
cross sections of the proton-$^{12}$C system.
The Glauber calculations fairly well reproduce
the elastic and inelastic scattering cross sections
from the ground state to the $2^+$ and $4^+$ states
up to the second cross section minima 
at incident energy from 50 to 800\,MeV.
Here we stress that the two parameters of the wave function
are determined only from the static structure information,
the rms radius and $B(E2)$. No adjustable parameter
is introduced in this reaction theory, which strengthens
the predictive power.

In order to compare the Glauber calculation with
the standard approximated methods,
we plot in Fig.~\ref{12Cdiff.fig}, the cross sections with the DWIA
obtained by Eqs.~(\ref{DWIA-1.eq}) and~(\ref{DWIA-2.eq}).
For the elastic scattering differential cross sections,
the standard OLA results given by Eq.~(\ref{OLA.eq}) are also plotted.
The OLA results are almost identical with the Glauber ones
up to the second dip of the elastic scattering differential cross sections.
The OLA takes into account most of contributions due to
the multiple-scattering processes in the proton-nucleus scattering.
We remark a recent interesting application of the OLA in which
the surface diffuseness of the nuclear density distribution
can be extracted from the proton-nucleus elastic scattering~\cite{Hatakeyama18}.
The standard OLA appears to be more efficient expansion
than that done in the DWIA as the DWIA results can only reproduce
the elastic scattering cross sections
at the forward angle up to the first dip.

The deviation between the Glauber and DWIA calculations
becomes more apparent in the inelastic scattering cross sections.
Though the DWIA calculations reproduces the cross sections around
the peaks, we see large deviation at the forward angles,
and at the backward angles with increase in the incident energy.
The deviation becomes drastic in
the inelastic scattering cross sections to the $4^+$ state.
This is because the one-step approximation made in the DWIA
is not sufficient to describe the whole inelastic processes,
whereas the present theory fully takes
into account the multiple-scattering or multistep
processes within the Glauber theory. 
We will address this matter in detail later in Sec.~\ref{inel.sec}.

\begin{figure}[!h]
\centering\includegraphics[width=3in]{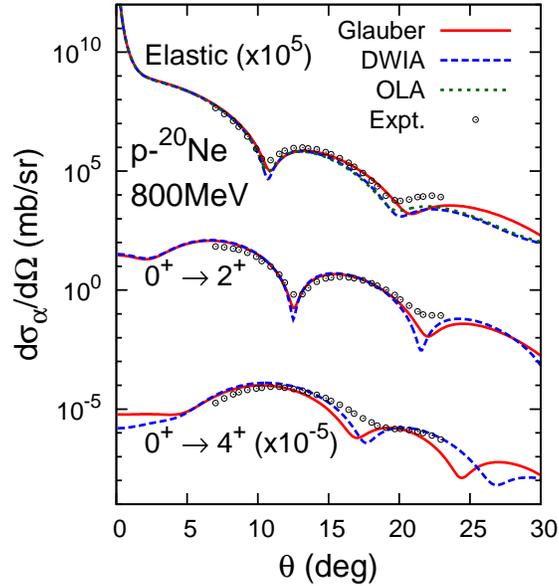}
\caption{Elastic and inelastic scattering differential cross sections
  of proton-$^{20}$Ne scattering at 800\,MeV.
  Experimental data are taken from Ref.~\cite{BRBFH88}.}
\label{20Nediff.fig}
\end{figure}

Figure~\ref{20Nediff.fig} displays
the elastic and inelastic scattering differential cross sections
of proton-$^{20}$Ne systems incident at 800\,MeV, where
the experimental data are available.
The theoretical calculations nicely reproduce the experimental cross
sections.
Though the difference between the Glauber and approximated
calculations is not as large as that of the proton-$^{12}$C case
in the elastic and $0^+\to 2^+$ inelastic scattering
differential cross sections,
we again see non-negligible difference in
the $0^+\to 4^+$ inelastic scattering cross sections.

\begin{figure}[!h]
\centering\includegraphics[width=\linewidth]{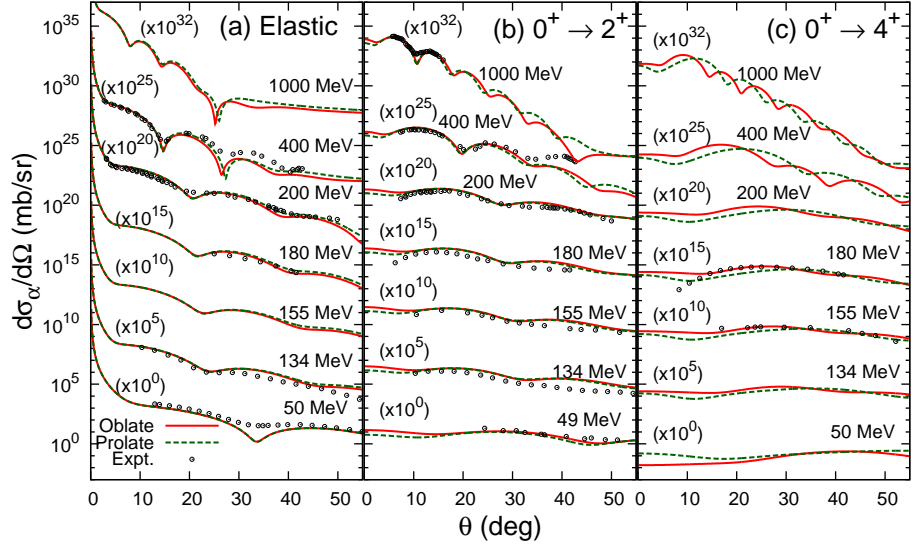}
\caption{(a) Elastic and (b) inelastic scattering differential cross sections
  of $0^+ \to 2^+$ and (c) $0^+ \to 4^+$
  for proton-$^{28}$Si scattering.
  The oblate and prolate wave functions are employed.
  See text for details. Experimental data are taken from
  Refs.~\cite{NSSOC83,HJLAJ88,CKSRJ90} for the elastic scattering,
  Refs.~\cite{AB84,HJLAJ88,CKSRJ90,ABVGD75} 
  for the $0^+\to 2^+$ inelastic scattering,
  and Refs.~\cite{AB84,CKSRJ90} for
    for the $0^+\to 4^+$ inelastic scattering.}
\label{28Sidiff.fig}
\end{figure}

Let us discuss proton-$^{28}$Si scattering,
where we consider both the oblate and prolate deformations
which cannot be constrained only by the $B(E2)$ value.
Figure~\ref{28Sidiff.fig} compares the elastic and
inelastic scattering differential cross sections with the prolate
and oblate wave functions of $^{28}$Si at incident energy
from 50\,MeV to 1\,GeV.
Again, overall agreement between the theory and experimental
cross sections is obtained.
For the elastic scattering differential cross sections,
the calculated cross sections with the oblate and prolate
wave functions give almost identical results because
the elastic scattering differential cross sections
at the forward angles are sensitive to the nuclear radius
which is taken as the same for the oblate and prolate wave functions
in this study.
For the inelastic scattering from the ground to the $2^+$ states,
we see small differences at the forward angles implying
that the cross sections have more information
about the quadrupole deformation than that of the $B(E2)$ value.
No difference between the cross sections
with the oblate and prolate wave functions is found
at the scattering angles where the experimental data are available.

A nuclear shape of $^{28}$Si has been attracted much interest for
a long time~\cite{Goodman70,Leander75,Bauhoff82}.
Recent microscopic model calculations showed
the oblate and prolate shapes coexist in
its spectrum~\cite{Enyo05,Ichikawa12,Chiba17}.
The difference between the oblate and prolate wave functions
can clearly be seen in the inelastic scattering differential cross
sections to the $4^+$ state.
The difference between the cross sections with
the oblate and prolate wave functions is significantly large at 155 and 180\,MeV
allowing one to distinguish the nuclear shape of $^{28}$Si.
The experimental cross sections are better reproduced by 
the theoretical cross sections with the oblate wave function.
We remark that the recent alpha-nucleus inelastic scattering measurement
supports the oblate ground state which is consistent with
the Skyrme-Hartree-Fock calculation with the SkM* interaction
~\cite{Peach16}.
Since the difference becomes more apparent at the higher incident energies,
measurement at such high energy will be important
as it reveals the nuclear shape of $^{28}$Si.
We note either the prolate or oblate shape
are assumed for $^{28}$Si wave function in this work.
Use of a more realistic wave function with a mixture of
the oblate and prolate shapes are interesting
to be worth studying in the future.

\section{Discussion: Incident energy dependence of the inelastic cross sections}
\label{inel.sec}

We have confirmed that our theory shows a fairly good description
of the inelastic scattering differential cross sections
to the $2^+$ and $4^+$ states
for $^{12}$C, $^{20}$Ne, $^{28}$Si in a wide range of the incident energies.
The magnitude of the inelastic scattering cross section to 
an angular momentum $J$ state is expected 
to be proportional to the $B(EJ)$ value.
In this section, we discuss which structure information
is actually probed by the inelastic cross sections
through an analysis of their incident energy dependence.

\begin{figure}[!h]
\centering\includegraphics[width=\linewidth]{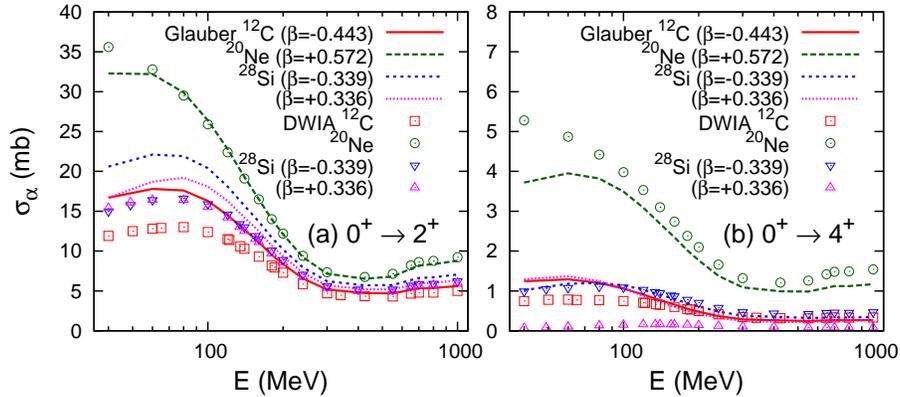}
\caption{Inelastic scattering cross sections
  of (a) $0^+ \to 2^+$ and (b) $0^+ \to 4^+$
    transitions for proton-$^{12}$C, $^{20}$Ne, and $^{28}$Si scattering
    as a function of incident energy.}
\label{inel.fig}
\end{figure}

Figure~\ref{inel.fig} displays the inelastic
scattering cross sections from the ground state
to the $2^+$ and $4^+$ states
for $^{12}$C, $^{20}$Ne, and $^{28}$Si as a function
of the incident energies.
The behavior follows the incident-energy dependence
of the $pN$ cross sections or the profile functions~\cite{Ibrahim08}:
The inelastic scattering cross sections
for all the nuclei are large at the low incident energies
and become smaller with increasing the incident energies
and again slightly increases at the higher energy end.
We see some difference between
the inelastic scattering cross sections
to the $2^+$ state with the oblate and prolate wave functions of $^{28}$Si
at the low incident energies
despite the fact that the two systems give
the same $B(E2)$ value, implying that
the proton-nucleus inelastic processes at low-incident energies
also contains the information other than that of the $B(E2)$ value
but some dynamical properties of the scattering.
The inelastic scattering cross sections to the $4^+$ state exhibit
the same trend and their magnitudes are one order of
magnitude smaller than those to the $2^+$ state.
The cross sections tend to be larger at the low-incident
energies where the effective interaction range becomes longer~\cite{Ibrahim08}
because the rotational excitation takes place
at the nuclear surface.

We also plot the results with the DWIA.
For the $0^+\to 2^+$ inelastic scattering,
the DWIA calculations work fairly well as the deviation from the Glauber
calculations are small at incident energies higher than $\sim$150\,MeV,
whereas they underestimate the Glauber 
cross sections at the lower incident energies.
For $^{20}$Ne, the DWIA calculations reproduce the Glauber
calculations even at the low-incident energies except for the lowest cases
where they are overestimated.
The effects of the multiple-scattering may be small
as the $^{20}$Ne nucleus is well deformed ($\beta_2=0.572$).

For the $0^+\to 4^+$ inelastic scattering,
the DWIA calculations show the similar incident-energy dependence
as we observed for the $0^+\to 2^+$ scattering
but for $^{12}$C case the DWIA underestimate and
for $^{20}$Ne case it overestimate the Glauber cross sections.
For $^{28}$Si, the cross sections with 
the oblate and prolate wave functions
show quite different behavior:
Despite the fact that the Glauber calculations with 
the oblate and prolate wave functions give the almost the same 
cross sections, the DWIA with the prolate wave 
functions predicts much smaller cross sections than
those with the oblate wave functions.
This trend may be related to the $B(E4; 4^+ \to 0^+)$ value
of $^{28}$Si: 1850 (50.4) $e^2$fm$^8$ with the oblate (prolate) wave
function, and $[B(E4)_{\rm oblate}/B(E4)_{\rm prolate}]^{1/4}\sim 2.5$.
Although a direct comparison between the $B(E4)$ value and
the $0^+ \to 4^+$ inelastic scattering cross section is not straightforward,
a smaller $B(E4)$ value gives a smaller inelastic cross section 
to the $4^+$ state with the DWIA calculation that
only takes into account the direct transition
from the ground state to the $4^+$ state.
However, in reality, the $0^+\to 4^+$ inelastic scattering 
occurs not only through the direct transition 
but also through the other multistep transitions leading to 
the same magnitude of the cross section with the oblate wave function.

\begin{figure}[!h]
\centering\includegraphics[width=\linewidth]{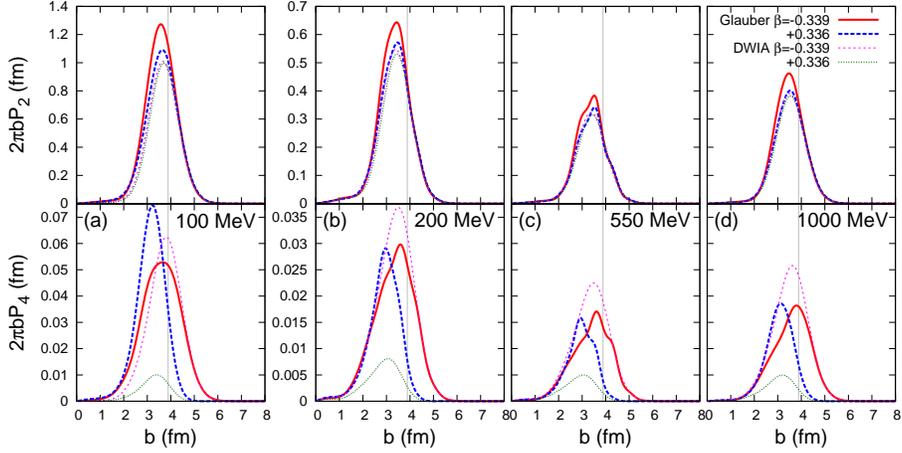}
  \caption{Reaction probabilities of
    the inelastic cross sections of (top) $0^+ \to 2^+$
    and (bottom) $0^+ \to 4^+$ transitions
    at incident energies of (a) 100, (b) 200, (c) 550, and (d) 1000\,MeV.
    The vertical thin line indicates
    the nuclear radius of $^{28}$Si,
    $\sqrt{\frac{5}{3}\left<r^2_p\right>}=3.89$\,fm.
    See text for details. }
  \label{prob.fig}
\end{figure}

The all discussions above become more transparent by calculating
the inelastic scattering reaction probability distribution defined by 
\begin{align}
P_{\alpha}(J_0\to J_\alpha;\bm{b})=\frac{1}{2J_0+1}\sum_{M_0,M_\alpha}
  \left|\mathcal{T}_{\alpha} (J_0M_0\to J_\alpha M_\alpha;\bm{b})\right|^2,
\label{prob.eq}
\end{align}
where $\sigma_\alpha=\int d\bm{b}\,P_{\alpha}(J_0\to J_\alpha;\bm{b})$.
Figure~\ref{prob.fig} plots
the reaction probability distributions defined by
Eq.~(\ref{prob.eq}) calculated with
the oblate and prolate wave functions of $^{28}$Si
as a function of impact parameters.
The incident energies are chosen as 100, 200, 550, and 1000\,MeV.
It can be clearly seen that the inelastic reaction mainly occurs
at the surface regions.
At the lower incident energies,
the probabilities show overall enhancement because the proton-nucleon
total cross sections become larger
and their effective interaction ranges are longer.
The probability distributions to the $2^+$ state
is similar to each other, whereas
these to the $4^+$ state behave quite differently
even their peak positions are different.
Recalling that the inelastic scattering differential cross sections
are obtained by a Fourier transform of the Glauber transition amplitude
$\sim \mathcal{T}_{\alpha} (J_0M_0\to J_\alpha M_\alpha;\bm{b})$,
the oblate and prolate natures of the wave function
is imprinted on the inelastic scattering differential cross
sections to the $4^+$ state as shown in Fig.~\ref{28Sidiff.fig}.

We also plot the results with the DWIA.
For the $2^+$ states, the DWIA works fairly well as we already see in
Fig.~\ref{inel.fig}.
In case of the $4^+$ state, the amplitudes with the DWIA calculations
are overestimated (underestimated) than the Glauber 
calculations for the oblate (prolate) wave functions, respectively,
and even their peak positions are different.
Since the multiple-scattering effects are significant
and mask the direct hexadecapole transition through
the inelastic scattering processes,
the $0^+\to 4^+$ inelastic scattering cross sections
cannot be a direct observable of $B(E4)$ in this incident energy range. 

\section{Application: Interaction cross sections}
\label{sigmai.sec}

So far, we have investigated the inelastic scattering cross sections
of $^{12}$C, $^{20}$Ne, and $^{28}$Si.
In this section, as an application of this theory,
we evaluate how large these inelastic scattering
cross sections in comparison to the total reaction cross section ($\sigma_R$).
This will be practically useful to compare 
with the experimentally observed interaction cross sections $(\sigma_I)$,
which can directly be measured by the transmission method 
as a change of the mass number~\cite{Tanihata85}
that involves the inelastic cross sections
to the bound excited states (BES).
Since one needs to make complicated corrections to obtain $\sigma_R$
experimentally, for a practical reason,
$\sigma_I\simeq \sigma_R$ has often been assumed.

Though it may be a good approximation at high incident energy
as all collisions lead to the direct breakup to the unbound states,
this difference actually affects the precision of the radius extraction.
The reliable estimation of the inelastic cross sections
is important, e.g., for the precise determination of the neutron-skin thickness
using the method proposed in Ref.~\cite{Horiuchi14,Horiuchi16}
in which the incident-energy dependence
of $\sigma_R$ on a proton target is utilized.
Though, in this work, the projectile nuclei is limited only to
$^{12}$C, $^{20}$Ne, and $^{28}$Si,
it is useful to know how much contributions of such inelastic processes
involved in $\sigma_R$.

We quantify the energy dependence of
the difference between $\sigma_I$ and $\sigma_R$.
Theoretically $\sigma_R$ can easily be calculated
by subtracting the survival probability $|\mathcal{T}_{0}(\bm{b})|^2$
from unity and integrating it over $\bm{b}$ as
\begin{align}
  \sigma_R=\int d\bm{b}\,\left(1-|\mathcal{T}_{0}(\bm{b})|^2\right).
\end{align}
To obtain $\sigma_I$,
one has to make corrections due to the inelastic processes
to the BES,
which make some complications to the theoretical calculations.
$\sigma_I$ can be evaluated by subtracting all
the inelastic cross sections going to the BES from $\sigma_R$ as
\begin{align}
  \sigma_I=\sigma_R-\sum_{\alpha\in {\rm BES}}\sigma_{\alpha},
\end{align}
where the BES included in
the calculations are the $2^+$ state for $^{12}$C;
and the $2^+$ and $4^+$ states for $^{20}$Ne and $^{28}$Si.
We note that $^{28}$Si has the other BES
which cannot be described by the rotational excitation.
Therefore, the $\sigma_I$ values for $^{28}$Si
evaluated in this paper provide their upper limit.

\begin{figure}[!h]
\centering\includegraphics[width=4in]{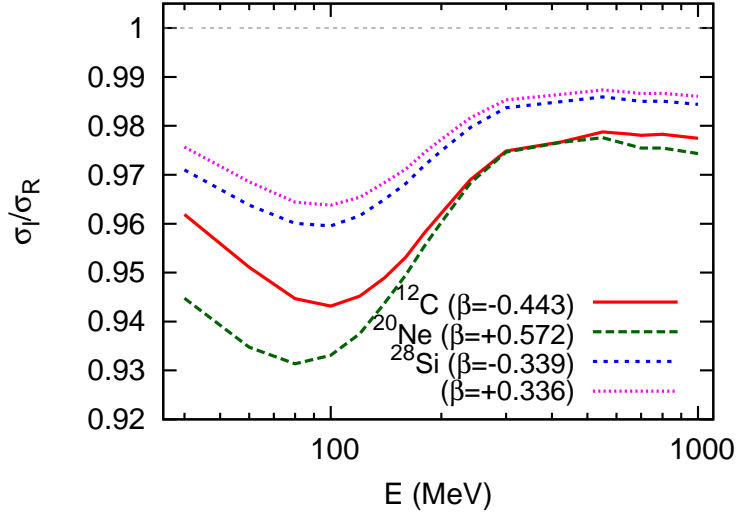}
  \caption{Ratio of the interaction 
    to total reaction cross sections, $\sigma_I/\sigma_R$,
    for proton-$^{12}$C, $^{20}$Ne,
    and $^{28}$Si scattering as a function of incident energy.
    Note that only $0^+ \to 2^+$ and $0^+ \to 4^+$ inelastic scattering cross sections
    are considered for $^{28}$Si, giving the upper limit of $\sigma_I$.
  See text for more details.}
\label{sigmaI.fig}
\end{figure}

Figure~\ref{sigmaI.fig} plots the ratio of the
interaction cross section to the total reaction
cross section, $\sigma_I/\sigma_R$. 
As expected, the difference between $\sigma_I$ and
$\sigma_R$ becomes small with increasing the incident energies,
which are 2--3\% to $\sigma_R$.
Since the inelastic reaction or rotational
excitation mainly occurs in the surface region
of the nucleus, the inelastic cross section increases 
and becomes at most $\sim 5$\% contribution to the total reaction
cross section at around 100\,MeV, where the effective interaction range
becomes longer than that at the high incident energy.
In the case of $^{20}$Ne, where the $B(E2)$ values are largest
among the four wave functions, the difference becomes largest
at most $\sim 7$\%,

One needs to care about those possible uncertainties in
the radius extraction using a proton probe.
For unstable nuclei, in general, the number of BES
is smaller than those of the stable nuclei.
Thus, the difference of $\sigma_I$ and $\sigma_R$
is expected to be less (or zero, if there is no BES)
than the cases presented in this paper.

\section{Summary}
\label{summary.sec}

In order to bridge the nuclear wave function with
the direct reaction observables,
we have performed a parameter-free reaction calculation for
the high-energy proton-nucleus inelastic processes
based on the Glauber theory.
The multiple-scattering processes within the Glauber theory
are fully taken into account by evaluating the Glauber amplitude
completely with the help of the factorization technique.
Inputs to the theory are the profile function
and the wave function of a target (or projectile) nucleus.
Once they are set, the theory has no adjustable parameter.

A power of this method has been demonstrated
by some examples of the proton inelastic scattering
of $^{12}$C, $^{20}$Ne, and $^{28}$Si.
Their ground- and excited-state wave functions
are described by the angular momentum projection
of a deformed intrinsic wave function.
The axially-symmetric harmonic-oscillator wave function
is used as a simplest choice where
its size and degree-of-deformation is fixed
so as to reproduce the measured charge radius and
reduced electric-quadrupole transition probability.
Experimental elastic and inelastic scattering differential cross section
data are well reproduced without introducing any adjustable parameter.
Our results show the higher order terms in the multiple-scattering operator
play a significant role in describing such inelastic scattering reactions
where the nuclear excitation occurs mainly at the surface region.
We find that the inelastic scattering differential cross sections
to the $4^+$ state of $^{28}$Si at the high incident energies
are useful observable to study the details
of the wave function, i.e., the nuclear shape.
We have shown that the multiple-scattering or multistep processes
are important in describing the $0^+ \to 4^+$
inelastic scattering cross sections,
which is not only through the direct $0^+\to 4^+$ transition 
but also through the other multistep transitions.
As an application of this theory,
we evaluate the contribution of the inelastic processes to the total reaction
cross section which can be useful to estimate the uncertainties in
the radius extraction from the interaction cross section measurement.

In this paper, we have tested the validity of our method
for well-known nuclei using a simple
deformed-harmonic-oscillator wave function.
The reaction theory developed in this work is rather general
formulation as it only requires a multi-Slater determinant wave function.
Use of elaborated wave functions is interesting that unveils the structure
of unstable nuclei and the role of the excess neutrons (protons)
with a systematic measurement of the inelastic scattering cross sections.
The extension along this direction is straightforward and
will be reported elsewhere. 

\section*{Acknowledgment}

We thank J. Singh for a careful reading of the manuscript.
This work was in part supported by JSPS KAKENHI Grant Numbers
18K03635 and 18H04569.

\appendix

\section{Evaluation of the scattering amplitude}
  
Let us write Eq.~(\ref{samp.eq}) in a more calculable form.
Since the beam is unpolarized,
we can choose any direction for the axis for the quantization.
Taking $z$ as the quantization axis,
the Glauber amplitude can easily be factorized as
\begin{align}
  \mathcal{T}_{\alpha} (J_0M_0\to J_\alpha M_\alpha;\bm{b})
  =\tau_\alpha(J_0M_0\to J_\alpha M_\alpha;b)e^{i(M_0-M_\alpha)\phi}.
\end{align}
With this expression,
we can easily carry out the integration over the azimuthal angle $\phi$
by expressing $\bm{b}$ with the polar coordinate $(b,\phi)$.
The scattering amplitude (\ref{samp.eq}) can be written more explicitly as
\begin{align}
  f_{\alpha}(q)=i^{|M_0-M_\alpha|+1}k\int_0^\infty db\,b
  J_{|M_0-M_\alpha|}(qb)\tau_\alpha(J_0M_0\to J_\alpha M_\alpha;b).
\end{align}
Here the integral expression of the Bessel function of the first kind
\begin{align}
J_n(x)=\frac{1}{2\pi i^n}\int_0^{2\pi}e^{ix\cos \phi+in\phi}\,d\phi, \qquad (n\geq 0),
\end{align}
is used. 

\section{Matrix elements with rotated wave functions}

As Eq.~(\ref{gaex.eq}) in Sec.~\ref{formulation.sec},
the angular momentum projected total wave function is expressed
by a superposition of many Slater determinant wave functions.
In this Appendix, we give more details about the evaluation
of the Glauber amplitude.
In this paper, we assume the ground- and excited-state wave functions
are generated from the same intrinsic state.
The expression of the amplitude with an $A$-body operator
$\prod_{j=1}^A\mathcal{O}_j$
can be written with the rotated single-particle wave functions as
\begin{align}
  &\left<\Phi_{\alpha; J_\alpha M_\alpha}\right|\prod_{j=1}^A\mathcal{O}_j\left|\Phi_{0; J_0M_0}\right>\notag\\
  &=\mathcal{N}_{M_\alpha 0}^{J_\alpha}\mathcal{N}_{M_00}^{J_0}
  \iint\,d\omega^\prime\,d\omega\,
  [\mathcal{D}_{M_\alpha 0}^{J_\alpha}(\omega^\prime)]^*\mathcal{D}_{M_0 0}^{J_0}(\omega)\,{\rm det}\left\{\mathcal{M}(\omega^\prime,\omega)\right\}
\end{align}
with
\begin{align}
  \left\{\mathcal{M}(\omega^\prime,\omega)\right\}_{jl}=
\int d\bm{r} [\phi_j(\bm{r}(\omega^\prime))]^*
  \mathcal{O}_j
  \phi_l(\bm{r}(\omega))
  \sum_{m^\prime,m} [\mathcal{D}^{1/2}_{m_j m^\prime}(\omega^\prime)]^*\mathcal{D}^{1/2}_{m_l m}(\omega)\delta_{m_j,m_l},
\end{align}
where $\bm{r}(\omega)=R^{-1}(\omega)\bm{r}$
with the inverse of the rotation matrix $R(\omega)$.
Note that the single-particle wave function
between the other rotated states is not orthogonal. 
The matrix element of a one-body operator
$\sum_{j=1}^A\mathcal{O}_j$ can also be evaluated with
\begin{align}
  &\left<\Phi_{\alpha; J_\alpha M_\alpha}\right|
  \sum_{j=1}^A\mathcal{O}_j
  \left|\Phi_{0; J_0 M_0}\right>\notag\\
  &=\mathcal{N}_{M_\alpha 0}^{J_\alpha}\mathcal{N}_{M_0 0}^{J_0}  \sum_{j,l=1}^A
  \iint\,d\omega^\prime\,d\omega\,
   [\mathcal{D}_{M_\alpha 0}^{J_\alpha}(\omega^\prime)]^*\mathcal{D}_{M_00}^{J_0}(\omega)\,
\{\mathcal{M}(\omega^\prime,\omega)\}_{jl}\,
  {\rm det}\left\{\tilde{\mathcal{B}}^{(jl)}(\omega^\prime,\omega)\right\},
\end{align}
where $\tilde{\mathcal{B}}^{(jl)}$ is a cofactor matrix
obtained by omitting the $j$th row and the $l$th column
from a matrix $\mathcal{B}$
whose elements are defined by
\begin{align}
  \left\{\mathcal{B}(\omega^\prime,\omega)\right\}_{st}=\int\,d\bm{r}\,[\phi_s(\bm{r}(\omega^\prime))]^*\phi_t(\bm{r}(\omega)).
\end{align}

\end{document}